\documentclass[conference]{IEEEtran}
\usepackage{balance} 
\usepackage{adjustbox, multirow,booktabs}
\usepackage{graphicx,caption,tikz} 
\usepackage{mathptmx, bm, amsfonts, amsmath,amsthm,breqn, listings}
\newtheorem{lemma}{Lemma}
\usepackage{url}
\usepackage{float,rotating} 
\usepackage{cite,hyperref}

\usepackage[linesnumbered,ruled,vlined]{algorithm2e}

\title{Towards a Decentralized IoT Onboarding for Smart Homes Using Consortium Blockchain}

\author{
    Narges Dadkhah, Khan Reaz, Gerhard Wunder\\
    Department of Mathematics and Computer Science\\ Freie Universität Berlin, Germany\\
    \{narges.dadkhah, g.wunder\}@fu-berlin.de, khanreaz@ieee.org
}

\makeatletter
\def\ps@IEEEtitlepagestyle{%
  \def\@oddfoot{\mycopyrightnotice}%
  \def\@oddhead{\hbox{}\@IEEEheaderstyle\leftmark\hfil\thepage}\relax
  \def\@evenhead{\@IEEEheaderstyle\thepage\hfil\leftmark\hbox{}}\relax
  \def\@evenfoot{}%
}
\def\mycopyrightnotice{%
  \begin{minipage}{\textwidth}
  \centering \scriptsize
   Copyright~\copyright~2025 IEEE. Personal use of this material is permitted. Permission from IEEE must be obtained for all other uses. The final version of record will be published in the 16th International Conference on Ubiquitous and Future Networks (ICUFN), 2025. The final published version will be available on IEEE Xplore.

  \end{minipage}
}
\makeatother
\begin{document}
\onecolumn 
\maketitle

\begin{abstract}

The increasing adoption of smart home devices and IoT-based security systems presents significant opportunities to enhance convenience, safety, and risk management for homeowners and service providers. However, secure onboarding—provisioning credentials and establishing trust with cloud platforms—remains a considerable challenge. Traditional onboarding methods often rely on centralized Public Key Infrastructure (PKI) models and manufacturer-controlled keys, which introduce security risks and limit the user's digital sovereignty. These limitations hinder the widespread deployment of scalable IoT solutions. This paper presents a novel onboarding framework that builds upon existing network-layer onboarding techniques and extends them to the application layer to address these challenges. By integrating consortium blockchain technology, we propose a decentralized onboarding mechanism that enhances transparency, security, and monitoring for smart home architectures. The architecture supports device registration, key revocation, access control management, and risk detection through event-driven alerts across dedicated blockchain channels and smart contracts. To evaluate the framework, we formally model the protocol using the Tamarin Prover under the Dolev-Yao adversary model. The analysis focuses on authentication, token integrity, key confidentiality, and resilience over public channels. A prototype implementation demonstrates the system's viability in smart home settings, with verification completing in 0.34 seconds, highlighting its scalability and suitability for constrained devices and diverse stakeholders. Additionally, performance evaluation shows that the blockchain-based approach effectively handles varying workloads, maintains high throughput and low latency, and supports near real-time IoT data processing.

\end{abstract}

\section{Introduction}

The proliferation of smart home devices, such as connected cameras and environmental sensors, has enhanced convenience and security for modern homeowners. At the same time, insurance providers increasingly encourage customers to install IoT-based security systems to improve risk mitigation and coverage. According to the Digital Market Insights report \cite{statista_smart_home_report}, the number of active households with smart-home security products is projected to reach 1.1 billion by 2029, with household penetration increasing from 42.8\% in 2024 to an estimated 51.1\% by 2029.
However, despite the anticipated benefits, security concerns continue to hinder IoT adoption. In modern hybrid warfare, critical infrastructure is increasingly targeted by both physical attacks and cyber operations. In their 2024 Digital Defense Report, Microsoft has reported a surge in attacks on poorly secured, internet-connected devices that control vital real-world processes~\cite{microsoft_digital_defense_report_2024}.

Generally, IoT onboarding consists of two distinct stages: (1) \emph{Network-layer onboarding}, establishing a secure network connection; (2) \emph{Application-layer onboarding}, configuring the device to securely communicate with cloud applications. A fundamental challenge in IoT deployment is secure and seamless device onboarding—the process of provisioning credentials and establishing a trusted link between a device and its target cloud or platform. This onboarding step often introduces significant friction for end-users, limiting widespread adoption. While multiple companies and working groups have proposed various onboarding solutions~\cite{fido2021,intelsdo,oath2reference}, no universal industry-wide standard has been widely adopted yet. Recognizing this challenge, NIST launched an initiative in 2020 to explore network onboarding solutions~\cite{nistnetwork-onboarding}.

Existing onboarding approaches typically assume that the end customer is known at the time of manufacture, requiring devices to be pre-configured during silicon fabrication. Additionally, many solutions rely on discrete secure elements for credential storage. The Achilles' heel of this method lies in the trust model: manufacturers hold control over root key generation, introducing security risks. Recent reports on backdoors in critical infrastructure have raised global concerns about supply chain security, as state-backed coercion could compel manufacturers to compromise root keys or install backdoors for espionage purposes. Furthermore, traditional Public Key Infrastructure (PKI)-based models restrict users' control over their device identities, as key management is often centralized across multiple infrastructure layers.

Centralized solutions pose concerns in the event of server failures or cyberattacks, as they may become inaccessible. In the case of a server attack, adversaries can only obtain the public keys, which do not pose an immediate security threat unless they gain access to the corresponding private keys stored in IoT devices. However, the process of regenerating the PKI introduces additional complexity, operational overhead, and inconvenience, further complicating recovery efforts.
To address these challenges, there is a growing need for decentralized human-in-the-loop onboarding mechanisms that eliminate dependence on device manufacturers' PKI and closed-source software.

This research builds upon the ComPass~\cite{reazComPassIMIS2021} and ASOP~\cite{reaz:asop-2022} protocols, which utilize wireless channel reciprocity to enable secure network-layer onboarding through strong symmetric key derivation, thereby avoiding reliance on public key cryptography and vendor-locked solutions. In this paper, we extend the previous protocols by introducing a tamper-proof, decentralized authentication mechanism that mitigates single points of failure. This extension addresses prior authentication challenges by integrating private permissioned blockchain technology. Beyond identity verification, our approach enforces robust access control mechanisms, ensuring that IoT data is accessible only to authenticated users and services with the appropriate permissions. Additionally, we take a step forward by developing a risk management solution that informs relevant organizations within the blockchain ecosystem, enabling timely detection and appropriate corrective action. To evaluate the proposed solution, we formally model the protocol using the Tamarin Prover~\cite{tamarin2013,tamarin-git} under the Dolev-Yao adversary model~\cite{dolev1983security}. The evaluation focuses on analyzing security properties such as \emph{authentication, key confidentiality, mutual authentication}, and \emph{public channel resilience}. The model allows for the simulation of adversarial behaviors such as message interception and manipulation to assess the protocol’s resilience against potential attack scenarios. Moreover, our blockchain solution was evaluated to confirm its effectiveness in managing IoT data in real time. Latency and throughput were analyzed under varying transaction arrival rates to assess the system’s performance across diverse load conditions.

The remainder of the paper is organized as follows: Section~\ref{R-Work} reviews related work. Section~\ref{preliminaries} provides background and introduces the proposed solution. Section~\ref{Implementation} outlines the system architecture and implementation. Section~\ref{Evaluation} presents evaluation results and security analysis. Finally, Section~\ref{conclusion} summarizes key findings and future directions.

\section{Related Work} \label{R-Work}
Ensuring security in smart home environments has become a critical challenge due to the vast amount of sensitive personal data collected by IoT devices, including information about daily habits, health conditions, and home activities. Unauthorized access to these devices poses serious risks to privacy and safety. Client impersonation, where attackers masquerade as legitimate devices to gain unauthorized access, has been identified as a key attack vector in~\cite{baucas2021iot}, highlighting the urgent need for robust authentication mechanisms.

To address these challenges, various authentication approaches have been proposed. The Fast Identity Online (FIDO) protocol~\cite{fido2021} is widely used for passwordless authentication, leveraging public key cryptography, where a private key remains securely stored on the user's device, and the corresponding public key is registered with a remote server.
While FIDO offers strong protection against phishing, its reliance on a dedicated \emph{Rendezvous Server} creates a single point of failure.
A breach at this server could allow an attacker to tamper with registered public keys. Additionally, if the private key is compromised—either through malware or physical access to the device—the entire authentication scheme is rendered ineffective. 
\begin{figure}[htbp]
    \centering
    \includegraphics[ width=8.5 cm, height=5.5 cm]{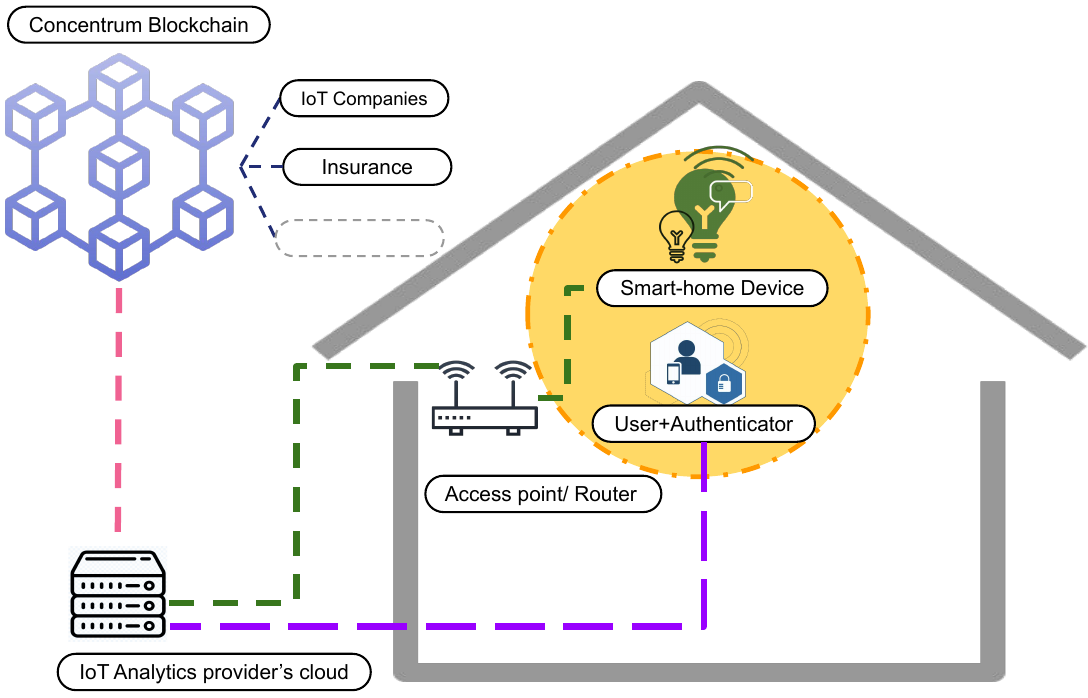}
    \caption{General architecture of IoT-based smart home}
     \label{fig:Architecture}
\end{figure}

The \emph{Matter} protocol, developed by the Connectivity Standards Alliance, offers a unified application-layer standard for smart home devices~\cite{matter_specification}. Matter's commissioning process leverages manufacturer-provisioned Device Attestation Certificates combined with Password-Authenticated Key Exchange (PAKE) for secure onboarding. All subsequent communications rely on certificate-based authenticated sessions. Matter's seamless integration with ecosystems such as Apple Home, Google Home, and Amazon Alexa enhances interoperability, but the requirement for persistent cryptographic key storage and certificate management poses challenges for resource-constrained devices.
Amazon's Frustration-Free Setup (FFS)~\cite{amazon_ffs} offers a proprietary, cloud-dependent onboarding process where devices purchased through Amazon can automatically receive Wi-Fi credentials via nearby Alexa devices. While FFS significantly improves user convenience, its reliance on Amazon's centralized infrastructure raises concerns over privacy and interoperability with non-Amazon ecosystems.

To mitigate these risks, researchers have proposed integrating a blockchain solution with FIDO-based authentication. Blockchain, as a decentralized and tamper-resistant ledger, addresses the single point of failure in FIDO systems. Studies such as~\cite{ou2024decentralized,gong2021blockchain} propose storing public keys on the blockchain rather than on central servers, leveraging blockchain's immutability to prevent unauthorized key modifications. However, such solutions still rely on FIDO's core mechanisms, including the \emph{Rendezvous server}, which remains vulnerable to large-scale Denial-of-Service (DoS) attacks.
Other research explores using IoT devices themselves as blockchain nodes, enabling devices to directly validate and record transactions~\cite{arif2020investigating}. This decentralized model reduces reliance on central authorities and enhances transparency in device interactions. However, practical challenges remain, particularly for resource-constrained devices which often lack sufficient processing power, storage, and energy to participate in computationally intensive blockchain consensus protocols. 

\section{Preliminaries and Proposed Method} 
\label{preliminaries}
\begin{figure}[htbp]
    \centering
    \includegraphics[ width=8.5 cm, height=5.2 cm]
{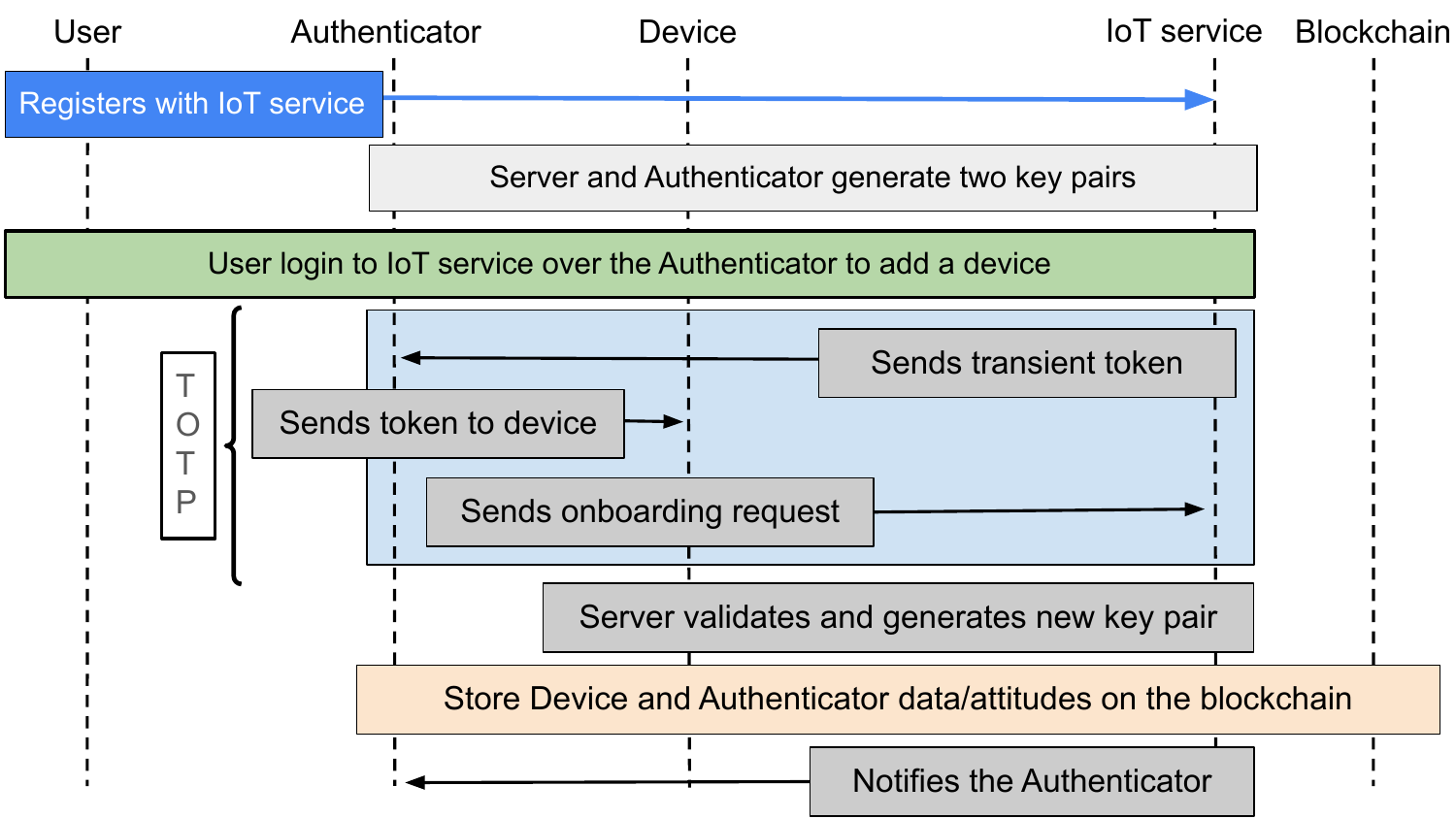}
    \caption{Sequence diagram of the proposed method}
    \label{fig:ASOP-flow}
\end{figure}

\subsection{Consortium Blockchain} 

Since the introduction of blockchain technology by Satoshi Nakamoto in 2008~\cite{nakamoto2008bitcoin}, its applications have expanded beyond financial transactions to a wide range of industries, including smart home systems. In this research, we employ Hyperledger Fabric (HLF)~\cite{androulaki2018hyperledger}, a well-known consortium blockchain. 
In HLF, only authenticated and approved participants can join the network, ensuring controlled access. Each participant is issued a public-private key pair and a certificate by the Certificate Authority (CA), establishing verifiable identities. Identity management is handled by the Membership Service Provider (MSP), which maps participants' cryptographic identities to network roles recognized across the blockchain. Clients initiate transactions, which are processed using HLF's execute-order-validate consensus mechanism. This ensures transactions are first executed, then ordered, and finally validated by a majority of nodes before being recorded on the blockchain. All communication within the network is encrypted and authenticated using TLS, ensuring data confidentiality and integrity. HLF's support for multiple channels allows for the creation of isolated ledgers, ensuring data are only accessible to authorized members of each channel, enhancing both privacy and security. Smart contracts (SC), referred to as chaincode, are deployed on designated channels within the network. They define the business logic, enabling automated and secure transaction processing with customizable access controls.

\subsection{TLS 1.3 with Pre-Shared Keys (PSK) Mode}
TLS 1.3 introduces the ability to establish secure communication using Pre-Shared Keys (PSKs), offering a lower-latency handshake compared to traditional certificate-based authentication~\cite{rfc8446}. The handshake in PSK mode eliminates the need for costly public key operations and significantly reduces the round trip time (RTT), making it well-suited for constrained environments like IoT devices. The PSK is combined with Ephemeral Diffie-Hellman (DHE) key exchange (PSK-DHE mode) to provide Perfect Forward Secrecy (PFS), ensuring that the compromise of the PSK does not compromise past session keys.

\subsection{Entities}

A generalized smart home architecture is presented in Fig.~\ref{fig:Architecture}, where the following entities are present:
\subsubsection{User $(U)$}
The user, often referred to as the end-user, is the individual who owns and operates the IoT device after purchasing it from a retailer or reseller. In the context of smart homes and small businesses, the user is typically a human capable of interacting with electronic systems, such as smartphones or tablets. However, this work does not consider large-scale industrial IoT deployments, where users may be automated systems or machine entities rather than individuals.

\subsubsection{Device $(D)$}
IoT devices are manufactured in large-scale production facilities, commonly operated by Original Equipment Manufacturers (OEMs). These manufacturers produce components and assemble devices on behalf of various companies. During silicon fabrication, the OEM must establish a root-of-trust identity for each device, typically in the form of a universally unique identifier (UUID). Depending on the device type, the manufacturer may also generate a root certificate based on public key cryptography, which is then anchored to the OEM’s certificate chain for authentication purposes.

\subsubsection{Authenticator $(A)$}
An authenticator refers to a handheld smart device, such as a smartphone or tablet, that facilitates secure communication and authentication. These devices are equipped with a rich user interface and support multiple wireless communication technologies, including Wi-Fi, cellular networks, and Bluetooth. The authenticator is primarily user-operated, requiring authentication through mechanisms such as passwords, PINs, biometric recognition (e.g., fingerprint or facial authentication), or other access control methods before it can be used.

\subsubsection{Server $(S)$}
A server represents the computational infrastructure that powers an IoT analytics provider’s services, either hosted in a cloud environment or deployed on-premises. It also serves as a management hub, enabling communication with the blockchain to store IoT information. The server may exist as a physical or virtual instance and is accessible via the internet or, in cases of local deployment, within a private network. To ensure security, server APIs are protected through authentication and authorization mechanisms, restricting access only to authorized entities.

\subsubsection{Hyperledger Fabric $(HLF)$}
\begin{figure}[htbp]
    \centering
    \includegraphics[width=8.5 cm, height=5 cm]{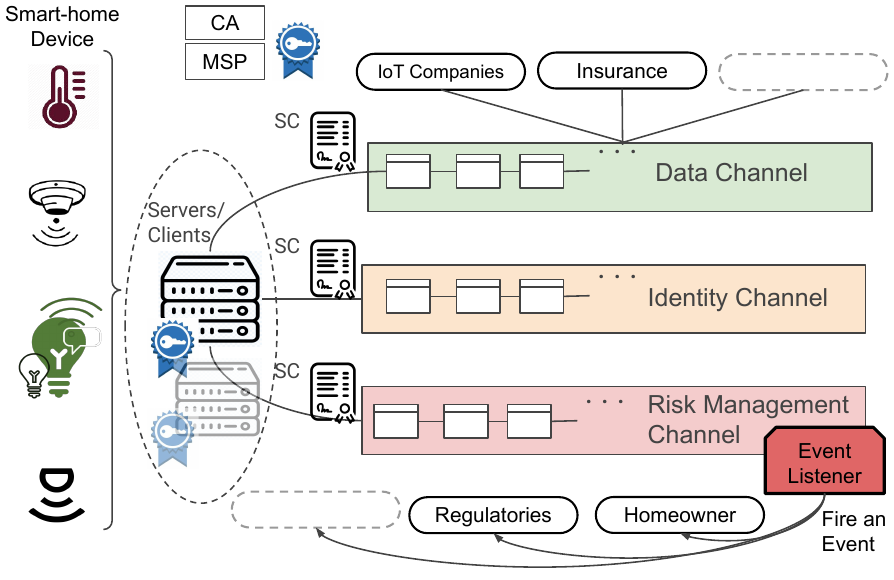}
    \caption{Interaction between HLF and smart home entities}
    \label{fig:HLF}
\end{figure}

In our approach, the server transmits all IoT transactions to the blockchain network and stores them securely. 
The server, in conjunction with the CA and MSP in the HLF network, is assigned the role of a client application and granted the necessary permissions to submit transactions within the HLF. In this system, three separate channels for different purposes are established: \textit{Identity Channel, Data Channel, and Risk Management Channel}. As shown in Fig. \ref{fig:HLF}, each channel is governed by its own dedicated SC, enabling the secure storage of data tailored to specific purposes while ensuring that distinct access control policies are enforced.

\subsection{Entity Interactions}
Algorithm \ref{alg:asop-Pseudocode} details the complete sequence of intermediate interactions in this approach. 
\subsubsection{Device Registration and Token Exchange}

A typical onboarding scenario begins when a user purchases an IoT device, such as a smart IP camera, to enhance home security and obtain potential insurance incentives. To use AI-powered surveillance services, the user selects an IoT analytics provider and installs the corresponding mobile application on a smartphone or tablet. This device, along with the application, functions as the \textit{Authenticator}. The user then registers an account with the IoT analytics provider, gaining access to the \textit{server's API}, which facilitates device enrollment, updates, and service utilization as shown in Fig. \ref{fig:ASOP-flow}.

\subsubsection{Secure Key Exchange and Session Establishment}
During account login, both the \textit{Server} and the \textit{Authenticator} generate 512-bits ephemeral key pairs based on the Module-Lattice-Based Key-Encapsulation Mechanism Standard~\cite{FIPS203}. These keys are denoted as $(S^{A}_{s}, S^{A}_{p})$ for the server (secret and public keys respectively) and $(A^{S}_{s}, A^{S}_{p})$ for the authenticator. Only public keys are exchanged, while each entity keeps its private key secret at all times. To limit long-term exposure, these keys remain valid for a short duration (for instance, hours or days), after which they are replaced through a chained renewal process. If keys expire without use, the \textit{Authenticator} prompts the user to re-login, triggering fresh key generation.

To protect against replay attacks and confirm mutual identity, the \textit{Server} sends a nonce $N_S$ to the \textit{Authenticator}. The \textit{Authenticator} signs $N_S$ with its private key $A^{S}_{s}$, producing $\{N_S\}_{\text{sign}(A^{S}_{s})}$. It further encrypts this signature with the server's public key $S^{A}_{p}$ before returning it. The \textit{Server} then decrypts and verifies the signature with $A^{S}_{p}$. For mutual authentication, the \textit{Server}  similarly signs a fresh nonce from the \textit{Authenticator} using its private key $S^{A}_{s}$. All communication is also protected by TLS 1.3 in a post-quantum hybrid mode (ML-KEM combined with ECDH)~\cite{cloudflareGooglePQC2020}, providing an additional layer of transport security.

When a user wants to register a new device, the \textit{Authenticator} sends a request to the \textit{Server}. The \textit{Server} generates a one-time token $T_n$ according to the TOTP algorithm~\cite{rfc6238}, with a  validity window of 30 seconds to lower the brute-force risk. 
The \textit{Server} encrypts $\{T_n, S_a\}$—where $S_a$ is its API address for the \textit{device}—using the authenticator's public key $A^{S}_{p}$ and sends it  to the \textit{Authenticator}. The \textit{Authenticator} decrypts the token and API address, appends the server's public key, and then re-encrypts with the device's public key $D^{S}_{p}$. The \textit{device}, which generates its own key pair specifically for the \textit{Server} $(D^{S}_{s}, D^{S}_{p})$, receives the token and attaches its pseudo-UUID $D_u$. It then encrypts data with the server's public key $S^{A}_{p}$ and sends it to the \textit{Server},  including a signature from the \textit{Authenticator} to prove token integrity. The \textit{Server} decrypts the message, verifies all signatures, and checks the token's TOTP validity to prevent replay.

\subsubsection{Device Activation and Secure Communication Setup}

Once the server confirms the device request, it creates a long-lived authentication token $T_D$, and generates a dedicated server key pair $(S^{D}_{s}, S^{D}_{p})$ for that device. It encrypts $\{T_D, S^{D}_{p}\}$ with $D^{S}_{p}$ so only the device can decrypt this information.
Instead of storing $\{T_D, S^{D}_{p}\}$ in the database, the server registers the device in the HLF. The server generates a transaction that includes the device information, ${T_D, S^{D}_{p}}$, along with \( A^{S}_{p} \), $D^{S}_{p}$, $D_u$, and the device status as \textit{active} and sends it to the HLF network. This transaction is stored in the \textit{Identity Channel}, allowing the server to monitor and manage active IoT devices within the smart home. Additionally, this approach provides an enhanced level of security, ensuring that each device is correctly registered with the appropriate authenticator. The system also serves as a reliable proof of ownership. If device ownership changes, a new record will be created in the HLF to reflect the current owner. With the help of SC, access to this channel remains restricted to servers only.  Simultaneously, the \textit{Server} notifies the \textit{Authenticator} that the device has been successfully connected by sending $\{D_u \,\|\, \langle \text{connected}\rangle\}$ encrypted with $A^{S}_{p}$. At this point, the device and the server each possess private keys and are aware of each other's public keys, ensuring they can authenticate one another, maintain confidentiality, and preserve message integrity in future exchanges.

\subsubsection{Storing Device Information} Once the device is registered, it can communicate directly with the server to transmit its collected data. Depending on the type of device, these data vary. The server receives these data and stores them, along with $D^{S}_{p}$, in the HLF \textit{Data Channel}. A key advantage of this approach is that, with the help of SC, it is possible to define different access control roles and allow manufacturers to access relevant information from their own devices. This is particularly useful for notifying the homeowner in cases such as when device warranties expire or when battery replacements are needed. Moreover, in the event of a home accident, the insurance company can access the data stored in this channel to determine whether the issue arose from human negligence or a device malfunction. Due to the tamperproof nature of blockchain, all data stored in these channels can also serve as a home's reputation record, potentially increasing its value for future home-related business. Additionally, the server can utilize SC to analyze device data for any anomalies, such as unusually high room temperatures or potential health risks. If such irregularities are detected by SC, the information is stored in the \textit{Risk Management Channel}. Our architecture
employs HLF channel-based event-triggering schemes in this channel,
enabling it to trigger alerts and notify relevant organizations, such as fire departments or ambulance services, in emergency situations. In this smart home solution, homeowners can also customize emergency contacts within the system, ensuring timely notifications in critical scenarios.
\subsubsection{Device De-registration and Key Revocation}
For de-registration or key revocation, the \textit{Authenticator} sends a revocation request to the \textit{Server} by encrypting $\langle \text{revoke}, D_u \rangle$ with $A^{S}_{p}$. The \textit{Server} then updates its records to invalidate the device's authentication token $T_D$ and adds the device's key to its Certificate Revocation List (CRL)~\cite{rfc5280} or marks it revoked via Online Certificate Status Protocol (OCSP) stapling~\cite{rfc6960}. It also sends a new transaction to the \textit{Device Channel} in HLF, including all relevant information from the registration phase, but with the device status set to ``Deactivated".

\begin{algorithm}[t]
    \small
    \caption{\small Pseudocode of the protocol steps}
    \label{alg:asop-Pseudocode}
    \DontPrintSemicolon
    \SetAlgoLined
    \SetNoFillComment
    \text{A:}$(S^A_s, S^A_p) \gets \text{ML-KEM}()$,    
    \text{S:}$(A^S_s, A^S_p) \gets \text{ML-KEM}()$ \\
    $T_n \gets \text{TOTP}()$,  $S_a \gets \text{API\_address}$ \\
    $encrypted\_data \gets \text{Encrypt}(A^S_p, (T_n, S_a))$ \\
    \textbf{Send} $S \to A: encrypted\_data$ \\

    $(T_n, S_a) \gets \text{Decrypt}(A^S_s, encrypted\_data)$ \\

    $encrypted\_token \gets \text{Encrypt}(S^A_p, T_n)$ \\
    $signature \gets \text{Sign}(A^S_s, encrypted\_token)$ \\

    $device\_data \gets C_K(S_a, S^A_p, signature)$ \\
    \textbf{Send} $A \to D: device\_data$ \\

    $(D^S_s, D^S_p) \gets \text{GenerateKeyPair}()$ \\
    $D_u \gets \text{GeneratePseudoUUID}()$ \\
    $message \gets D^S_p \parallel D_u \parallel signature$ \\
    $encrypted\_message \gets \text{Encrypt}(S^A_p, message)$ \\
    \textbf{Send} $D \to S: encrypted\_message$ \\

    $message \gets \text{Decrypt}(S^A_s, encrypted\_message)$\\
    $(D^S_p, D_u, signature) \gets \text{Parse}(message)$ \\
    $isValid \gets \text{VerifySignature}(A^S_p, encrypted\_token, signature) \land \text{CheckTOTPValidity}(T_n)$ \\

    \If{$isValid$}{
        \textbf{RegisterDevice}$(D_u, D^S_p)$ \\
        $T_D \gets \text{GenerateLongLivedToken}()$ \\
        $(S^D_s, S^D_p) \gets \text{GenerateKeyPair}()$ \\
        $device\_response \gets \text{Encrypt}(D^S_p, (T_D, S^D_p))$ \\
        
        \textbf{Send} $S \to HLF: CreateTrx(T_D, S^{D}_{p}, D^{S}_{p}, A^{S}_{p}, D_u, \textit{active}) $ \\
        
        \textbf{Send} $S \to D: device\_response$ \\
        $auth\_notification \gets \text{Encrypt}(A^S_p, (D_u \parallel \texttt{<connected>}))$ \\
        \textbf{Send} $S \to A: auth\_notification$ \\
    }

\end{algorithm}

\section{Implementation} \label{Implementation}

To demonstrate the feasibility of our onboarding framework, we integrated it with the IoTree backend platform~\cite{iotree_platform}, a commercial LoRaWAN-based device and network management service. We assume users have valid login credentials for IoTree. The onboarding process begins with the user launching the Authenticator app to log into the IoTree web portal. After successful authentication, the user initiates device registration by submitting a join request. In response, the IoTree backend returns a JSON payload containing a time-sensitive token generated via the TOTP algorithm and an API endpoint URL.  Once provisioned via the ComPass protocol, the IoT device receives the API endpoint from the Authenticator app and issues a POST request with the populated fields.  We deployed a consortium blockchain using HLF, featuring Node.js smart contracts, three channels, and an event service. The system runs on Debian 11 with 16 GB RAM and an Intel i7 CPU, with all components containerized via Docker. An Android application, developed in Kotlin, compatible with API level 30 (Android 11) and above, serves as the Authenticator.

\section{Evaluation} \label{Evaluation}

\subsection{Formal Verification using Tamarin Prover }
The Tamarin Prover is a state-of-the-art tool for formally verifying the security properties of cryptographic protocols~\cite{tamarin-git}. It models protocol rules, message flows, and adversary capabilities symbolically, then exhaustively explores all possible traces under the Dolev-Yao adversary model to verify whether specified security properties (lemmas) hold.

The proposed method defines two distinct communication paths among the \emph{Authenticator}, \emph{Device}, and \emph{Server}. The channel between the \emph{Authenticator} and \emph{Server}, denoted as $H_s$, is assumed to be secure and authenticated, established via a TLS 1.3 handshake. In contrast, the channel between the \emph{Device} and \emph{Server}, $H_p$, is public and unauthenticated, offering no inherent trust or security and exposing communications to potential adversaries. To model the protocol accurately in Tamarin, these channels must reflect their security characteristics.

To validate the core security properties of our protocol, we define formal lemmas in the Tamarin Prover. These capture guarantees such as correct device registration, token integrity and binding, and resistance to token reuse or tampering. We present three key lemmas addressing authentication, token integrity, and keypair confidentiality—essential for establishing trust in the onboarding process.

\begin{lemma}[\emph{Authentication}: The registration success message can only be sent if the device's request was validated]
 
\begin{equation*}
\begin{aligned}
    \forall \, \text{token} \text{deviceID} \text{nonce} \#i \#j. \,
     (\text{In}(\text{sign}(\text{validate}(\langle \text{token}, \text{nonce} \rangle),
     \text{device\_key}))) 
     @ \#i \land \text{Out}(\text{registration\_success}
    \\
    @ (\text{deviceID}))  \#j ) \Rightarrow (\#i < \#j)
\end{aligned}
\end{equation*}
\end{lemma}
\begin{lemma}[\emph{Token Integrity}: A transient token can only be used by the device that received it]

\begin{equation*}
\begin{aligned}
    \forall \, \text{token} \text{deviceID1}\text{deviceID2} \#i \#j. \,
     ((\text{Out}(\text{sign}(\text{device\_request} 
      (\text{deviceID1}, \text{token}, \text{nonce}), \text{auth\_key}))
    @ \#i) 
    \land \text{In}
      (\text{sign}\\(\text{device\_request} 
      (\text{deviceID2},  \text{token}, \text{nonce}), \text{auth\_key})) 
    @ \#j ))
      \Rightarrow (\text{deviceID1} = \text{deviceID2})
\end{aligned}
\end{equation*}

  \end{lemma}

\begin{lemma}[\emph{Keypair Confidentiality}: The key pair generated for the device is confidential]

\begin{equation*}
\begin{aligned}
    \forall \, \text{new\_keypair\_device} \#i \#j. \,
     (( \text{Out}(\text{new\_keypair} (\text{new\_keypair\_device}))  
      @ \#i ) 
    \land \text{K}(\text{new\_keypair\_device})  
    @ \#j ) ) 
    \Rightarrow (\#i = \#j)
\end{aligned}
\end{equation*}

\end{lemma}

\begin{table}[htbp]
    \centering
     \caption{Tamarin output summary after evaluation}
    \begin{tabular}{@{}l l@{}}
        \hline
        \textbf{Analyzed:} & ASOP-protocol.spthy \\
        \hline
        Authentication (all-traces): & verified (4 steps) \\
        Token\_Integrity (all-traces): & verified (4 steps) \\
        Keypair\_Confidentiality (all-traces): & verified (8 steps) \\
        \hline
    \end{tabular}
   
    \label{table:asop-tamarin}
\end{table}

Table~\ref{table:asop-tamarin} summarizes that all lemmas were validated across all execution traces, including adversary interactions. The results confirm that authentication, token integrity, and key confidentiality properties hold under the defined threat model, with verification completed in 0.34 seconds.

\subsection{Transaction Latancy}

\begin{figure}[htbp]
    \centering
    \includegraphics[width=7 cm, height=4 cm]{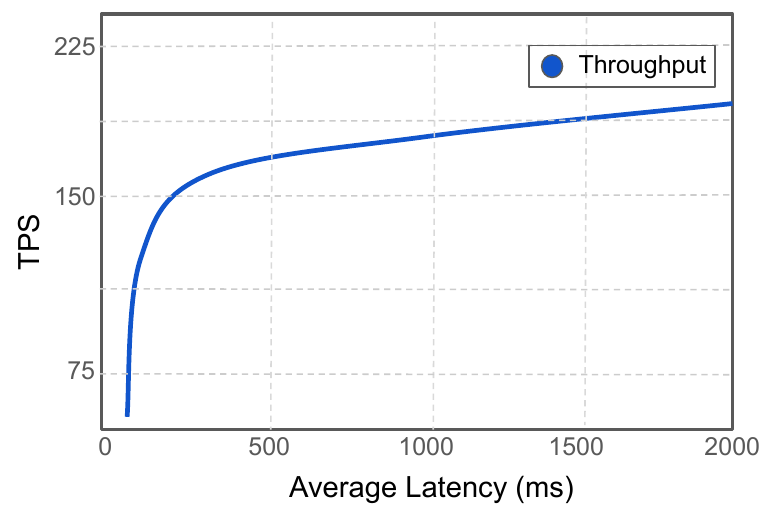}
    \caption{Latency vs Throughput performance}
    \label{fig:HLF-result}
\end{figure}

To evaluate transaction throughput in the HLF system, tests were conducted across input rates ranging from 30 to 300 transactions per second (TPS). For each load level, the number of successfully processed transactions and average latency were recorded. Results show that the system maintains consistent throughput and low latency under moderate workloads. In particular, latency remains below 500 milliseconds up to approximately 175 TPS as shown in Fig.~\ref{fig:HLF-result}. Beyond this point, latency increases noticeably, growing rapidly with higher transaction rates. These results indicate that while the system performs efficiently under mid-level traffic, it approaches saturation at higher loads, suggesting the need for optimization or infrastructure improvements to sustain performance.

\section{conclusion} \label{conclusion}

This paper presented a decentralized onboarding framework tailored for IoT devices. It combines symmetric-key authentication with blockchain-based access control, leveraging HLF for device registration, data reporting, and risk monitoring. The architecture defines separate trust paths among the entities using dedicated HLF channels to enforce fine-grained access control. SCs manage key operations including registration, revocation, and anomaly detection, while channel-based events enable real-time alerts. Security properties—authentication, token integrity, key confidentiality, and public channel resilience—were formally verified using the Tamarin Prover under the Dolev-Yao model, confirming protocol soundness and efficiency. Performance evaluation showed stable throughput and low latency under moderate loads, with latency increasing at higher traffic due to resource limitations, indicating the need for further optimization.

\section*{Acknowledgements}

This work is supported by the Federal Ministry of Education and Research (BMBF) of Germany through the projects 6G-RIC (16KISK025), UltraSec (16KIS1682), and PHY2APP (16KIS1473).

\balance
\bibliographystyle{IEEEtran}
\bibliography{main.bib}
\end{document}